\documentclass[sigconf]{acmart}

\AtBeginDocument{%
  \providecommand\BibTeX{{%
    \normalfont B\kern-0.5em{\scshape i\kern-0.25em b}\kern-0.8em\TeX}}}

\copyrightyear{2025}
\acmYear{2025}
\setcopyright{acmlicensed}\acmConference[CIKM '25]{Proceedings of the 34th ACM International Conference on Information and Knowledge Management}{November 10--14, 2025}{Seoul, Republic of Korea}
\acmBooktitle{Proceedings of the 34th ACM International Conference on Information and Knowledge Management (CIKM '25), November 10--14, 2025, Seoul, Republic of Korea}
\acmDOI{10.1145/3746252.3761538}
\acmISBN{979-8-4007-2040-6/2025/11}

\usepackage{multirow}
\usepackage{multicol}
\usepackage{scalefnt}

\usepackage{tikz}
\usepackage{pgfplots}
\usepackage{float}
\usepackage{algorithm}
\usepackage{algorithmic}
\usepackage{graphicx}
\usepackage{subcaption}

\begin{document}

\title{Personalized Tree-Based Progressive Regression Model for Watch-Time Prediction in Short Video Recommendation}

\author{Xiaokai Chen}
\authornote{Both authors contributed equally to this research.}
\email{dzhchxk@126.com}
\affiliation{%
  \institution{Kuaishou Technology}
  \city{Beijing}
  \country{China}
}

\author{Xiao Lin}
\authornotemark[1]
\email{jackielinxiao@gmail.com}
\affiliation{%
  \institution{Kuaishou Technology}
  \city{Beijing}
  \country{China}
}

\author{Changcheng Li}
\email{lichangcheng@kuaishou.com}
\affiliation{%
  \institution{Kuaishou Technology}
  \city{Beijing}
  \country{China}
}

\author{Peng Jiang}
\email{jiangpeng@kuaishou.com}
\affiliation{%
  \institution{Kuaishou Technology}
  \city{Beijing}
  \country{China}
}

\renewcommand{\shortauthors}{Xiaokai Chen, Xiao Lin, Changcheng Li, and Peng Jiang}

\begin{abstract}
In online video platforms, accurate watch time prediction has become a fundamental and challenging problem in video recommendation. Previous research has revealed that the accuracy of watch time prediction highly depends on both the transformation of watch-time labels and the decomposition of the estimation process. TPM (Tree based Progressive Regression Model) achieves State-of-the-Art performance with a carefully designed and effective decomposition paradigm. TPM discretizes the watch time into several ordinal intervals and organizes them into a binary decision tree, where each node corresponds to a specific interval. At each non-leaf node, a binary classifier is used to determine the specific interval in which the watch time variable most likely falls, based on the prediction outcome at its parent node.

The tree structure is central to TPM, as it defines the decomposition of watch time estimation and how ordinal intervals are discretized. However, TPM uses a predefined full binary tree, which may be sub-optimal for two reasons. First, full binary trees imply equal partitioning of the watch time space, which may fail to capture the complexity of real-world distributions. Second, rather than relying on a fixed global structure, we advocate for a personalized, data-driven tree that can be learned end-to-end. Thus, we propose PTPM to enable highly personalized decomposition of watch estimation with better efficacy and efficiency. Moreover, we show that TPM suffers from selection bias due to conditional modeling and propose a simple solution. We conduct extensive experiments on offline datasets and online environments. Offline results show improved watch time accuracy, and online A/B tests further validate the effectiveness of our framework. PTPM has been fully deployed in core traffic scenarios and now serves over 400 million users daily.
\end{abstract}





\maketitle

\section{Introduction}
In video-sharing platforms (e.g., TikTok and Kuaishou), watch time denotes
the amount of time users spend on watching videos, which is strongly related to user engagement and Daily Active Users (DAU). Watch time prediction is thus an essential task in online video recommender systems. 

Previous studies \cite{lin2023tree,sun2024cread} have highlighted several challenges in watch time prediction: (1) Value regression such as MSE treats overestimation and underestimation equally, ignoring the ordinal relationships among predictions, which are essential for producing accurate rankings in RecSys. (2) Conditional dependence exists in video viewing behavior and should be integrated into models; for example, one has to watch the first half of the video before he/she finishes watching the whole one. (3) For reliable watch time predictions, models should be aware of prediction uncertainty. Although existing work \cite{covington2016deep,zhan2022deconfounding,zhang2023leveraging,sun2024cread} tackle some important limitations, none of them have fully considered these issues.

\begin{figure}[htbp]
    \centering
    \includegraphics[width=0.4\textwidth]{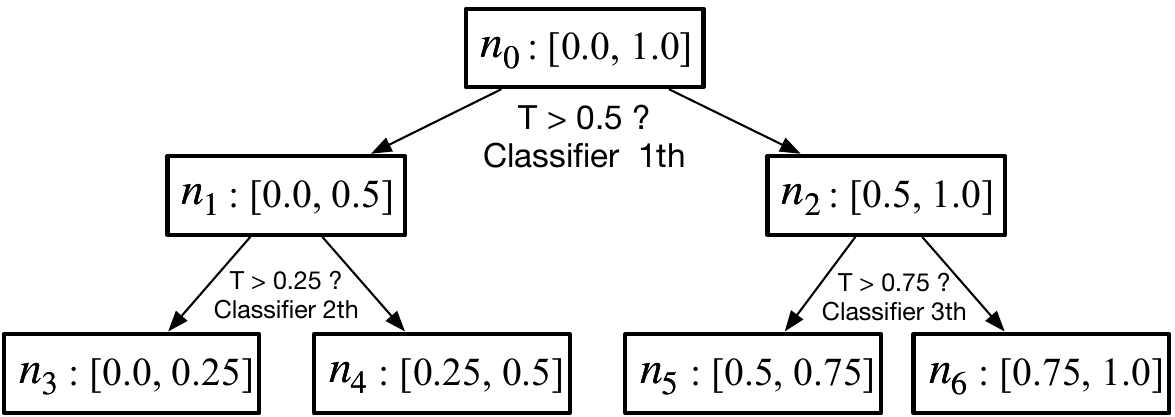}
    \caption{Schematic of TPM. The watch time regression task is decomposed into tree-structured classification tasks.}
    \label{fig1}
\end{figure}

\vspace{-7pt}
The Tree based Progressive Regression Model (TPM) \cite{lin2023tree} is a representative work aimed at addressing the aforementioned challenges simultaneously. As shown in \textbf{Fig \ref{fig1}}, TPM divides the range of watch time into ordinal intervals, constructing a binary tree where each node signifies an interval. This approach converts the watch time regression task into a series of binary classification tasks. In the tree, each non-leaf node is assigned with a binary task to predict the correct child node that includes the watch time label. Each edge represents a possible result of the prediction and leads to the next node. (1) TPM captures ordinal relationships through tree-structured ordinal classification tasks, where binary ordinal labels are constructed by comparing the watch time with a threshold. (2) TPM models tree-structured conditional dependency, where each child task decides which interval watch time falls into given the outcome of its parent task. (3) TPM can estimate the distribution of watch time by traversing from root to leaf, thereby quantifying uncertainty, i.e., the variance of the distribution.

\begin{figure}[t]
    \centering
\begin{subfigure}[b]{0.45\columnwidth}
    \begin{tikzpicture}[scale=0.65]
\begin{axis}[
    ybar,
    symbolic x coords={
        0,
        0.0333,
        0.0666,
        0.0999,
        0.1332,
        0.1665,
        0.1998,
        0.2331,
        0.2664,
        0.2997,
        0.333,
        0.3663,
        0.3996,
        0.4329,
        0.4662,
        0.4995,
        0.5328,
        0.5661,
        0.5994,
        0.6327,
        0.666,
        0.6993,
        0.7326,
        0.7659,
        0.7992,
        0.8325,
        0.8658,
        0.8991,
        0.9324,
        0.9657,
        0.999
    },
    xtick={0, 0.1998, 0.3996, 0.5994, 0.7992, 0.999},
    xticklabels={0, 0.2, 0.4, 0.6, 0.8, 1.0},
    x tick label style={rotate=0},
    xlabel={Avg. Watch Time per User (normalized)},
    ylabel={\# Sampled Users},
    ylabel style={yshift=-10pt}, 
    ymode=log,
    ymin=1,
    ymax=100000,
    ytick={1, 10, 100, 1000, 10000, 100000},
    yticklabels={1, 10, 100, 1k, 10k, 100k},
    ymajorgrids, 
    grid style={dashed, gray}, 
    width=6.5cm,
    height=5.0cm,
    bar width=4pt,
    axis x line*=bottom, 
    axis y line*=left,   
]
\addplot coordinates {
    (0, 70124)
    (0.0333, 20362)
    (0.0666, 5037)
    (0.0999, 1917)
    (0.1332, 870)
    (0.1665, 507)
    (0.1998, 348)
    (0.2331, 202)
    (0.2664, 151)
    (0.2997, 100)
    (0.333, 72)
    (0.3663, 52)
    (0.3996, 44)
    (0.4329, 30)
    (0.4662, 25)
    (0.4995, 30)
    (0.5328, 21)
    (0.5661, 17)
    (0.5994, 22)
    (0.6327, 11)
    (0.666, 11)
    (0.6993, 6)
    (0.7326, 7)
    (0.7659, 6)
    (0.7992, 2)
    (0.8325, 2)
    (0.8658, 3)
    (0.8991, 4)
    (0.9324, 3)
    (0.9657, 3)
    (0.999, 3)
};
\end{axis}
\end{tikzpicture}
\caption{User side}
\end{subfigure}
    \hspace{0.25cm} 
    \begin{subfigure}[b]{0.45\columnwidth}
        \begin{tikzpicture}[scale=0.65]
    
        \begin{axis}[
            xlabel={Watch time normalized by duration},
            ylabel={Quantile},
            ylabel style={yshift=-12pt}, 
            legend style={at={(0.95,0.05)}, anchor=south east, font=\small, nodes={scale=0.8, transform shape}},
            legend cell align={left},
            cycle list name=color list,
            grid=major,
            width=6.5cm,
            height=5cm,
            line width=0.5pt , 
            ymax=1.0,
            ymin=0, 
            xmin=0, 
            xmax=1.0
        ]
        
    \addplot+[mark=none,line width=0.7pt, smooth] coordinates {
        (0.012719934890043043,0.03125)(0.013303333948633613,0.0625)(0.014528471971673812,0.09375)(0.015538604498963778,0.125)(0.016604383129932655,0.15625)(0.017758690265523427,0.1875)(0.02457539222480814,0.21875)(0.030596884457479392,0.25)(0.03435759334877193,0.28125)(0.03811289630134191,0.3125)(0.04083894345727363,0.34375)(0.04150828099950299,0.375)(0.04920517771727666,0.40625)(0.057269839229419445,0.4375)(0.062120054388929515,0.46875)(0.07518261608553509,0.5)(0.09242324599642986,0.53125)(0.11482134765595466,0.5625)(0.15597915656562172,0.59375)(0.21073494663081563,0.625)(0.2913906542817567,0.65625)(0.37099999384640897,0.6875)(0.45048921666415165,0.71875)(0.5323968432593033,0.75)(0.6168971894186521,0.78125)(0.6963094791850769,0.8125)(0.712227795766492,0.84375)(0.8031592778793419,0.875)(0.9462892544665178,0.90625)(1.0081916375686915,0.9375)(1.0429412410343444,0.96875)
    };
    \addlegendentry{Music}

    \addplot+[mark=none,line width=0.7pt, smooth] coordinates {
        (0.013312923276741813,0.03125)(0.021075679562111514,0.0625)(0.02711909806786816,0.09375)(0.029690475136164372,0.125)(0.03220085077267565,0.15625)(0.03403823150115924,0.1875)(0.035556757656458056,0.21875)(0.03816866050660405,0.25)(0.04144676259328955,0.28125)(0.044222264747536134,0.3125)(0.04470334038666537,0.34375)(0.05709313161632369,0.375)(0.07217058433308811,0.40625)(0.08999085027969389,0.4375)(0.10773418170577728,0.46875)(0.11748466257668712,0.5)(0.15046090550300376,0.53125)(0.15956101859299687,0.5625)(0.17291207077955736,0.59375)(0.20962605124535025,0.625)(0.2799169181436647,0.65625)(0.38647879350571146,0.6875)(0.583439598790433,0.71875)(0.8195635359131058,0.75)(0.9824460141896794,0.78125)(1.0031335468894538,0.8125)(1.011272011948565,0.84375)(1.012540672451193,0.875)(1.020168598873268,0.90625)(1.1909728076396504,0.9375)(1.519883103096975,0.96875)
    };
    \addlegendentry{Game}
    
    \addplot+[mark=none,line width=0.7pt, smooth] coordinates {
            (0.006331845238095238,0.03125)(0.014028829597360092,0.0625)(0.01542887904387653,0.09375)(0.02219750889679715,0.125)(0.02657753311707739,0.15625)(0.027555209053618284,0.1875)(0.030693375224367575,0.21875)(0.03390640927117887,0.25)(0.03523533653846154,0.28125)(0.036122384865409266,0.3125)(0.03847078309090144,0.34375)(0.04316710915085856,0.375)(0.04375739028974696,0.40625)(0.04507642410126395,0.4375)(0.051173259271972756,0.46875)(0.051673702755709205,0.5)(0.05331056497145462,0.53125)(0.05459844559585492,0.5625)(0.0578821243523316,0.59375)(0.06022021179017313,0.625)(0.06273100361723535,0.65625)(0.06602331606217617,0.6875)(0.0676068652849741,0.71875)(0.07248704663212435,0.75)(0.09573721192249568,0.78125)(0.14146778949012878,0.8125)(0.1848404642779716,0.84375)(0.27536084943913663,0.875)(0.5921460700757576,0.90625)(1.0006065017899461,0.9375)(1.1591131573639895,0.96875)
        };
        \addlegendentry{Makeup}

    \addplot+[mark=none,line width=0.7pt, smooth] coordinates {
            (0.06476509393662253,0.03125)(0.06476509393662253,0.0625)(0.08831328289904719,0.09375)(0.11186147186147186,0.125)(0.11843073593073593,0.15625)(0.125,0.1875)(0.13447115384615385,0.21875)(0.1439423076923077,0.25)(0.2214539124668435,0.28125)(0.2989655172413793,0.3125)(0.30385579937304075,0.34375)(0.3087460815047022,0.375)(0.31365709946983755,0.40625)(0.318568117434973,0.4375)(0.455864145297573,0.46875)(0.5931601731601731,0.5)(0.7656709956709957,0.53125)(0.9381818181818182,0.5625)(0.9595631543270318,0.59375)(0.9809444904722452,0.625)(1.0151274176499157,0.65625)(1.0493103448275862,0.6875)(1.0973824451410659,0.71875)(1.1454545454545455,0.75)(1.371079062580479,0.78125)(1.5967035797064126,0.8125)(2.07181499331641,0.84375)(2.546926406926407,0.875)(2.7767532467532465,0.90625)(3.0065800865800867,0.9375)(3.387965367965368,0.96875)
        };
        \addlegendentry{Outfit}

    \addplot+[mark=none,line width=0.7pt, smooth] coordinates {
            (0.0026942435837371743,0.03125)(0.004430452518590663,0.0625)(0.006631346073734943,0.09375)(0.009135392275126237,0.125)(0.01222144135126522,0.15625)(0.01686699566786835,0.1875)(0.021625703238595667,0.21875)(0.026950193970928946,0.25)(0.033628386477752126,0.28125)(0.04108194078590379,0.3125)(0.049982128828883296,0.34375)(0.06002992410103272,0.375)(0.07238978278920778,0.40625)(0.09133084069637316,0.4375)(0.11240508011733923,0.46875)(0.13856230175225848,0.5)(0.17953476901894314,0.53125)(0.23477449091083474,0.5625)(0.30333599594369537,0.59375)(0.39831947680904334,0.625)(0.527291297519722,0.65625)(0.6779907502463329,0.6875)(0.8525877473324067,0.71875)(0.9848089732900575,0.75)(1.0060611510835973,0.78125)(1.0171912953678859,0.8125)(1.0273609917342328,0.84375)(1.0396567927241975,0.875)(1.061923715460316,0.90625)(1.1297135452276768,0.9375)(1.4104863404565324,0.96875)
        };
        \addlegendentry{Fun}

    \addplot+[mark=none,line width=0.7pt, smooth] coordinates {
            (0.03430950557162853,0.03125)(0.061325032930790814,0.0625)(0.08984557841552826,0.09375)(0.12149533347570364,0.125)(0.15020580032815517,0.15625)(0.17649094552522795,0.1875)(0.21609418319345278,0.21875)(0.2779712906065732,0.25)(0.35380490185734,0.28125)(0.445164106713625,0.3125)(0.6062962858312396,0.34375)(0.7592596531583977,0.375)(0.8597209049438966,0.40625)(0.9609482626664625,0.4375)(1.0129766523108437,0.46875)(1.0554788742420267,0.5)(1.0863515779023212,0.53125)(1.122088473772806,0.5625)(1.1574520359465252,0.59375)(1.2088993779985926,0.625)(1.2788783302502207,0.65625)(1.3927858614473345,0.6875)(1.527193935359004,0.71875)(1.7200566491421423,0.75)(1.8564808224008589,0.78125)(2.008408268316408,0.8125)(2.148218659115261,0.84375)(2.4122778779294363,0.875)(2.7604330716034022,0.90625)(3.308915096508959,0.9375)(4.428090750803613,0.96875)
        };
        \addlegendentry{Sports}

    \addplot+[mark=none,line width=0.7pt, smooth] coordinates {
            (0.0035189533713518954,0.03125)(0.0035189533713518954,0.0625)(0.007455809338674577,0.09375)(0.012967407692926331,0.125)(0.02851642008339083,0.15625)(0.057448651188805655,0.1875)(0.08506307339449541,0.21875)(0.10477064220183485,0.25)(0.1244782110091743,0.28125)(0.14536245099624134,0.3125)(0.16644280284659635,0.34375)(0.18845425979064787,0.375)(0.21116404555497179,0.40625)(0.2444235155002265,0.4375)(0.3040571958978083,0.46875)(0.36369087629539004,0.5)(0.4276564695699132,0.53125)(0.4916220628444363,0.5625)(0.5752219697444187,0.59375)(0.6666756020945849,0.625)(0.7744133798865265,0.65625)(0.9038633516008354,0.6875)(1.015304741283497,0.71875)(1.018694638776275,0.75)(1.0220845362690527,0.78125)(1.0299527605743282,0.8125)(1.0385673726816866,0.84375)(1.0472679064697572,0.875)(1.0560328815183622,0.90625)(1.2067201834862384,0.9375)(1.7122133027522934,0.96875)
        };
        \addlegendentry{Beauty}

        \end{axis}
    \end{tikzpicture}
    \caption{Item side}
    \end{subfigure}
    \vspace{-0.3cm}
    \caption{Watch time distributions differ significantly between the user and item sides on the Kuaishou platform. \textit{Left (a), user side}: Distribution of users' average watch time per view. \textit{Right (b), item side}: Cumulative distribution of watch time across different video categories.}
    \label{fig2}
\end{figure}
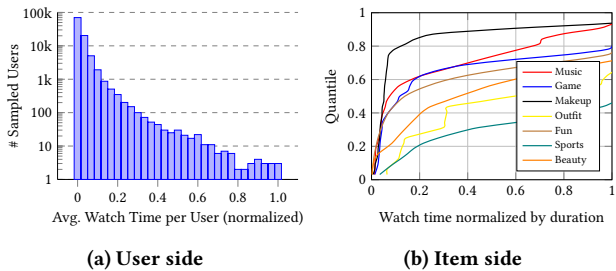

Despite its effectiveness, TPM has certain limitations. The default tree structure is a full binary tree, which governs both the decomposition of watch time prediction and the discretization of the watch time range. This design enforces equal partitioning of the watch time space. For example, the leaf intervals $n_3, n_4, n_5, n_6$ in \text{Fig \ref{fig1}} demonstrate such partitioning. Yet it reflects a rigid assumption that may fail to capture the complexity of real-world watch time distributions. As shown in \textbf{Fig \ref{fig2}}, watch time distributions vary substantially across users and items, making it infeasible to adopt a single discretization that fits all. Moreover, although TPM benefits from tree-based conditional modeling, it also inevitably suffers from sample selection bias that is inherently induced by the same mechanism. During training, each classifier is exposed only to samples with watch time values falling within a specific interval. However, during inference, the classifier may encounter samples associated with arbitrary ground-truth watch time values.

To address this limitation, we seek to develop a context-aware tree structure. This is crucial because learning a dynamic tree structure inherently involves learning an appropriate discretization of the watch time range, which is a key factor that allows the model to adapt to complex watch time distributions.
Despite its importance, the problem of learning such a discretization remains largely underexplored in the literature. Among discretization-based methods \cite{lin2023tree,zhang2023leveraging,sun2024cread}, only CREAD \cite{sun2024cread} attempts to study the discretization process. It adopts a heuristic formula to define discretization and uses grid search to optimize the choice between equal-width and equal-frequency binning strategies. However, CREAD suffers from three key limitations:
(1) its static, sample-invariant partitioning strategy compromises adaptability;
(2) its heuristic-constrained search space may overlook better solutions; and
(3) its inefficient search algorithm relies on grid search and requires retraining the classifier at each iteration.

In this paper, we propose PTPM to learn context-sensitive tree structure at the sample level (i.e., user-item pair) efficiently. PTPM introduces a Tree Structure Learning (TSL) module to optimize the tree structure via pruning. This design is motivated by the observation that a larger global tree naturally encompasses a variety of candidate subtrees, from which new structures can be efficiently derived through selective pruning. For instance, one can prune a full binary tree to obtain an unbalanced subtree. More importantly, the pruning paradigm enables efficient joint training of the tree structure and its associated classifiers. A pruning model is designed to output a binary vector, where each element indicates the pruning action for a corresponding node. The model accepts arbitrary user- and item-related features as input, thereby enabling personalized and dynamic tree structures. To train the model, we propose a self-critical objective that encourages the pruned subtree to outperform the original global tree. To address the aforementioned distribution shift between training and inference, PTPM further introduces a module that enables unbiased conditional learning.

The main contributions of this paper are as follows.
\vspace{-4pt}
\begin{itemize}
\item To the best of our knowledge, this is the first work to learn watch time discretization in an end-to-end manner, enabling the model to adapt to diverse user and item contexts and make more accurate predictions.
\item We propose PTPM, which extends TPM by learning context-aware tree structures through a pruning-based approach. This design enables sample-specific discretization of watch time by jointly optimizing the tree and its classifiers.
\item PTPM supports real-time personalized tree generation per request, while maintaining computational efficiency comparable to standard TPM. This is crucial for industrial recommender systems with strict latency requirements.
\item PTPM is fully deployed in Kuaishou's core recommendation services, delivering real-time recommendations to over 400 million daily active users. Extensive offline and online evaluations show significant performance improvements.
\end{itemize}

\vspace{-10pt}
\section{Related Work}

\textbf{Watch Time Prediction}
Watch time prediction estimates the dwell time for a candidate short video, a crucial task in industrial RecSys. Traditional value regression minimizes absolute prediction errors (i.e., MAE) but struggles with distribution imbalances. Weighted Logistic Regression \cite{covington2016deep} (WLR) uses watch time as a logistic sample weight to estimate watch time through click-through rate odds, but it exhibits systemic errors and is not suitable for auto-playing short videos. D2Q and DML \cite{zhan2022deconfounding,zhang2023leveraging} categorize videos by duration, mapping watch time to group-wise quantiles, and then learn to regress these quantiles for robust estimation. However, these methods primarily optimize for absolute prediction values, overlooking the ordinal nature of watch time in the ranking process of RecSys.

\textbf{Ordinal Regression}
Ordinal regression addresses problems where target labels have a natural order, such as age estimation \cite{chang2011ordinal} and search relevance prediction \cite{wu2003practical}, and is often implemented by converting ordinal targets into binary classification tasks. Following this idea, CREAD \cite{sun2024cread} defines each classifier to indicate whether the label exceeds a threshold. TPM \cite{lin2023tree} organizes ordinal intervals into a tree structure, where watch time prediction is formulated as traversing root-to-leaf paths.

A major challenge in ordinal regression is determining appropriate thresholds, an issue that remains underexplored. Among existing studies, only CREAD \cite{sun2024cread} explicitly addresses the discretization of continuous watch time. It uses a hand-crafted heuristic formula to determine thresholds and tunes hyperparameters via grid search over equal-width and equal-frequency schemes. However, CREAD has several limitations, including fixed, non-adaptive discretization, a constrained heuristic space, and the high computational cost of grid search due to repeated training.
In contrast, our proposed PTPM jointly trains the tree structure and its associated classifiers to learn a context-aware partitioning strategy.

\textbf{Tree Structure Learning}
Traditional tree-based models like XGBoost \cite{chen2016xgboost} use greedy algorithms to construct trees that partition the feature space. In contrast, our method divides the task space for adaptive watch time decomposition—a fundamental difference.
JTM \cite{zhu2019joint}, an extension of TDM \cite{zhu2018learning}, jointly learns the tree structure and user representations. PTPM differs in three key ways:
(1) It uses the tree for task decomposition and watch time estimation, while JTM focuses on item retrieval;
(2) It enables end-to-end tree learning, avoiding JTM’s greedy, layer-wise construction;
(3) It generates input-specific trees on the fly, unlike JTM’s static, shared structure.

\section{Preliminary}
Before presenting the details of PTPM, we begin by introducing the watch time prediction task and then provide a brief overview of the previous TPM approach. The notations is included in Table \ref{tab:1}.

\subsection{Watch Time Prediction Task}
Let $\{(  x_i,y_i)\}$ denote the user-item historical interactions, where $x_i \in \mathbb{R}^d$ denotes the $i$-th input vector, comprising user and video features, and $y_i\in \mathcal{Y} \subset \mathbb{R}^+$ is the ground truth watch time. The target of watch time prediction is to learn a model $f_\theta(\cdot)$ so that the prediction $\hat{y_i}=f_\theta(x_i)$ is close to $y_i$ under certain metrics.

\begin{table}
  \caption{Notations}
  \label{tab:1}
  \begin{tabular}{ccl}
    \toprule
    Notation & Meaning\\
    \midrule
    $y_i$ & Watch time label\\
    $\hat y_i$ & Predicted watch time\\
    $  x_i$ & Features of a user-video pair\\
    $T$ & Tree in TPM framework\\
    $N_{T}:\{n_i\}$ & set of nodes in ${T}$ \\
    $L_{T}:\{l_k\}$ & set of leaf nodes in ${T}$\\
    ${f}_{\theta_i}$ & the classifier assigned to $n_i$ with param. $\theta_i$\\
    $\phi_{l_k}$ & the path from root to leaf node $l_k$\\
    $n_{\phi_{l_k}(i)}$ & the node at level i along path $\phi_{l_k}$\\
    $d(l_k)$ & the depth of leaf node $l_k$\\
  \bottomrule
\end{tabular}
\end{table}

\subsection{The TPM Approach}
As shown in \textbf{Fig. \ref{fig1}}, the Tree-based Progressive Regression Model (TPM) \cite{lin2023tree} reformulates watch time prediction as a series of classification tasks over a tree $T$. The expected watch time is then estimated based on the predicted probabilities over the leaf intervals.

Specifically, the tree structure consists of the following components:
\textbf{Nodes:} Each node represents an interval, where the interval of a parent node is the union of those of its child nodes.  
\textbf{Classifiers:} Each non-leaf node is associated with a binary classifier that determines whether the watch time $y_i$ falls into the left or right child interval.  
\textbf{Edges:} Each edge corresponds to a prediction outcome and points to the subsequent node in the decision path.

\textbf{Calculating the probabilities $p(\hat y_i \in l_k \mid x_i, T)$ for all leaf intervals $l_k \in L_T$} involves traversing the decision path from the root to each leaf node. Formally, we follow the path $\phi_{l_k} = \{n_{\phi_{l_k}}(0), \ldots, n_{\phi_{l_k}}(d(l_k))\}$ from the root to the leaf node $ l_k $, where $ n_{\phi_{l_k}}(j) $ is the node at level $ j $, and $ d(l_k) $ is the depth of $ l_k $. Thus:
\begin{equation}\label{eq1}
\begin{aligned}
    p(\hat y_i\in l_k|  x_i, T) &= p(\hat y_i\in {n}_{\phi_{l_k}}(j), \forall j\leq d(l_k) |x_i, T)\\
    &=\prod_{1\leq j\leq d(l_k)}p(\hat y_i\in {n}_{\phi_{l_k}}(j) |\hat y_i\in {n}_{\phi_{l_k}}(j-1),x_i, T)
\end{aligned}
\end{equation}
 where $p(\hat y_i\in {n}_{\phi_{l_k}}(j) |\hat y_i\in {n}_{\phi_{l_k}}(j-1),x_i, T)$ is parameterized by the classifier ${f}_{{n}_{\phi_{l_k}}(j-1)}$ assigned to node ${n}_{\phi_{l_k}}(j-1)$.

 \textbf{Calculating the expected watch time} $\hat y_i$ can then be defined as follows:
 \begin{equation}\label{eq2}
    E(\hat y_i|x_i,{T}) = \sum_{l_k\in L_{{T}}} E(\hat y_i|\hat y_i\in l_k,x_i,{T}) p(\hat y_i\in l_k |x_i,{T})
\end{equation}
where the term $E(\hat y_i|\hat y_i\in l_k,x_i,{T})$ denotes the expected watch time for the leaf interval $l_k$, which, in the case of TPM, is approximated by the midpoint of that interval.

\textbf{Uncertainty modeling.} TPM predicts not only expected watch time but also the uncertainty of its predictions. Treating watch time as a random variable following a categorical distribution $p(\hat{y}_i \in l_k \mid x_i, T)$ for all $l_k \in L_{{T}}$, uncertainty is measured by the variance:
\begin{equation}\label{eq3}
    \text{Var}(\hat y_i|x_i,{T}) = E({\hat y_i}^2|x_i,{T}) - E(\hat 
 y_i|x_i,{T})^2
\end{equation}

\textbf{Training of TPM.} TPM comprises a series of classifiers, $\{f_{\theta_i}, i \in \{0, 1, \ldots, |N_T| - |L_T|\}\}$. Each classifier's objective is to determine the sub-interval, identified by its child nodes, to which the label $y_i$ belongs. For a sample $(x_i, y_i \in l_k)$, there is always a unique path  $\phi_{l_k}$ from the root to the leaf node $l_k$. This path determines the label for training the corresponding classifiers. In TPM, if $T$ is a binary tree, each classifier is binary, and samples belonging to the right-hand child node are considered positive samples.

The objective function of TPM consists of three components:
(1) Maximizing the likelihood w.r.t. $\mathcal{L}_{ce} = -\log(p(\hat{y_i} \in l_k | x_i, T))$, which is computed based on the classification errors of classifiers along the path $\phi_{l_k}$.
(2) Prediction variance: $\mathcal{L}_{var} = Var(\hat{y_i}|x_i,{T})$.
(3) Regression error: $\mathcal{L}_{reg}=\|E(\hat{y_i}|x_i,T) - y_i\|_2$.
The final objective function is:
\begin{equation}\label{eq4}
\mathop{\min}_\mathbf{\theta} \mathcal{L}_{TPM} =  \mathcal{L}_{ce}(\mathbf{\theta})  + \mathcal{L}_{reg}(\mathbf{\theta}) + \mathcal{L}_{var} (\mathbf{\theta})
\end{equation}
where $\mathbf{\theta}$ are the parameters of classifiers.
\subsection{Tree Construction in TPM}
In TPM, a full binary tree is predefined to ensure sample balance. Specifically, TPM calculates the global quantiles of watch time and discretizes watch time intervals based on these quantiles. The entire interval [0.0, 1.0] is initially assigned to the root node and is iteratively split into sub-intervals for non-root nodes. This division method ensures a balanced label distribution for each node, reducing learning difficulty \cite{johnson2019survey}. 

The full binary tree enforces an equal partitioning of the watch time range, which can be suboptimal for several reasons. First, equal partitioning fails to capture the diverse distribution patterns of watch time across different users or items (see Fig. \ref{fig2}). Second, from an optimization perspective, the tree structure can be treated as a learnable variable rather than being predefined (see Sec. \ref{sec41}).

\tikzset{global scale/.style={
    scale=#1,
    every node/.append style={scale=#1}
  }
}

\begin{figure}[t]
\captionsetup[subfigure]{font=footnotesize}
\centering
\scalefont{1.8}
\begin{tikzpicture}[semithick,->,grow=down, global scale = 0.5,level distance = 2.2cm,sibling distance=4.0cm]
\node [draw,rectangle, line width = 1pt] (root) {$n_0:[0.0,1.0]$}
child {node [draw,rectangle, line width = 1pt] (A) {$n_1:[0.0,0.25]$}}
child {
  node [draw,rectangle, line width = 1pt] (B) {$n_2:[0.25,1.0]$}
  child {node[draw, rectangle, line width = 1pt] (C) {$n_3:[0.25,0.5]$}}
  child {
    node[draw,rectangle, line width = 1pt] (D) {$n_4:[0.5,1.0]$}
  }
};

\begin{scope}[nodes={draw=none}]
  \path (root) -- (B) node [near start, right, xshift=5pt, yshift=-5pt] {$p(\hat y_i \in n_2 \mid n_0, T)$};
  \path (B) -- (D) node [near start, right, xshift=5pt, yshift=-5pt] {$p(\hat y_i \in n_4 \mid n_2, T)$};

  \path (root) -- (A) node [near start, left, yshift=-5pt] {$p(\hat y_i \in n_1 \mid n_0, T)$};
  \path (B) -- (C) node [near start, left, yshift=-5pt] {$p(\hat y_i \in n_3 \mid n_2, T)$};
\end{scope}

\end{tikzpicture}

\caption{The tree structure defines the number and targets of classifiers, and its leaf intervals ({$n_1,n_3,n4$}) concretely represent the discretization strategy.}

\label{fig:aaa}
\end{figure}
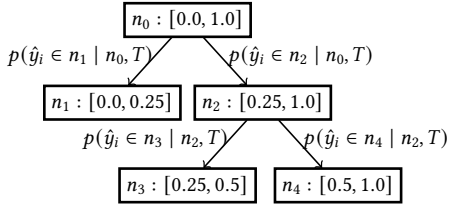

\section{PROPOSED METHOD}
We extend the original TPM framework by proposing PTPM, which integrates two additional modules: the Tree Structure Learning module (TSL) and the Unbiased Conditional Learning module (UCL). During the training phase, TSL enables joint learning of the tree structure and its classifiers. Meanwhile, the UCL module is designed to calibrate the biased classifiers. During inference, PTPM builds a context-aware tree for each request in real time.

\subsection{Tree Structure Learning} \label{sec41}

\subsubsection{Formulation}
As shown in \textbf{Fig \ref{fig:aaa}}, the tree structure determines the objective of the classifiers, indicating that learning a tree involves a bilevel optimization problem \cite{anandalingam1992hierarchical, colson2007overview}. Such a problem consists of two nested optimization tasks: an outer optimization whose objective depends on the solution of an inner optimization. Formally,
\begin{equation}\label{eq5}
\begin{aligned}
& \text{Outer:} \quad T^* = \arg\min_{T} \mathcal{L}_{\text{tree}}(\mathbf{\theta}^*(T), T) \\
& \text{Inner:} \quad \mathbf{\theta}^*(T) = \arg\min_{\boldsymbol{\theta}} \mathcal{L}_{\text{TPM}}(\boldsymbol{\theta}, T)
\end{aligned}
\end{equation}
where $T$ is the outer-level variable and $\mathbf{\theta}$ the inner-level one. The outer objective $\mathcal{L}_{tree}$ seeks the optimal tree $T^*$ to maximize watch time prediction performance (see Section \ref{sec412}), while the inner objective $\mathcal{L}_{\text{TPM}}$ aims to learn the optimal parameters $\boldsymbol{\theta}$ under a given tree structure $T$ (see Equation~\ref{eq4}).

The outer objective, $\mathcal{L}_{tree}(\mathbf{\theta}^*(T), T)$, depends on the inner solution $\mathbf{\theta}^*(T)$. Thus, optimizing Eq. \ref{eq5} is challenging, as $\mathbf{\theta}$ must converge for each $T$ before the outer process starts.

\begin{figure}[t]
    \centering
    \includegraphics[width=0.95\columnwidth]{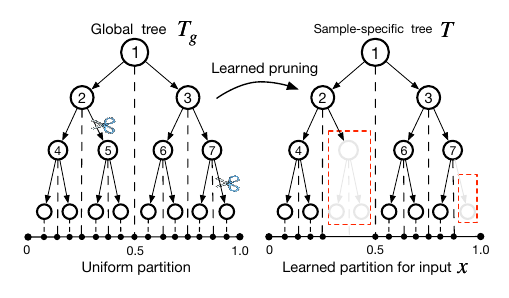}
    \caption{Illustration of the pruning process from a global tree $T_g$ to a pruned tree $T$. The global tree $T_g$ corresponds to a uniform partition, while the pruned tree $T$ defines a learned, non-uniform partition conditioned on sample $x$.}
    \label{fig3}
\end{figure}

To overcome this limitation, we propose the TSL module that decouples the training of the tree structure and the classifiers.
Specifically, we reformulate the tree structure learning task as a pruning problem: starting from a global tree with fully-trained classifiers, we generate a context-aware subtree for each sample pair $x_i$ through selective pruning, as illustrated in \textbf{Fig.}~\ref{fig3}. Importantly, pruning does
not alter the classifiers’ objectives but modifies the selection of
classifiers used to calculate the expected watch time, thus eliminating the costly inner optimization.

Formally, given the global tree $T_g$, the Eq. \ref{eq5} can be reformulated as follows:
\begin{equation}\label{eq6}
\begin{aligned}
& T^*=\mathop{\arg\min}_{T = T_g \circledast \mathbf{a}_T} \mathcal{L}_{tree}(\mathbf{\theta}^*(T), T) \\
\end{aligned}
\end{equation}
\begin{align}
\text{s.t.} \quad
{\mathbf \theta}^*(T) &=  {\mathbf \theta}^*(T_g) \odot (1-\mathbf{a}_T) \tag{a} \\
\mathbf{\theta}^*(T_g) &= \arg\min_\mathbf{\theta} \mathcal{L}_{TPM}(\mathbf{\theta}, T_g) \tag{b}
\end{align}
where $\mathbf{a}_T$ is a pruning vector, with each $a_i\in\{0,1\}$ indicates if the $i^{th}$ element should be pruned. The symbol $\odot$ signifies element-wise multiplication, and $\circledast$ denotes the pruning operation.
Condition (a) specifies that ${\mathbf \theta}^*(T)$ is efficiently derived from the globally trained ${\mathbf \theta}^*(T_g)$ via a pruning mask, eliminating the need for retraining during structure search.

A potential concern is that pruning the global model $T_g$ into different candidate structures $T$ might lead to a mismatch between the inherited parameters $\mathbf{\theta}^*(T)$ and the pruned architecture. However, we note that each $T$ is a strict subtree of $T_g$, and the pruning operation only disables certain branches without altering the semantics of the remaining paths. As such, the classification objective of any subtree $T$ remains consistent with its role in the full tree. Consequently, the inherited parameters $\mathbf{\theta}^*(T) = \mathbf{\theta}^*(T_g) \odot (1 - \mathbf{a}_T)$ remain well-adapted, and no additional fine-tuning is required to maintain predictive accuracy.

Eq. \ref{eq6} involves two variables: the classifier parameters $\mathbf{\theta}(T_g)$ of the global tree and the subtree $T$. For $\mathbf{\theta}(T_g)$, we train a vanilla TPM with the global tree. For $T$, we establish a pruning model to minimize $\mathcal{L}_{tree}$ for better watch time prediction accuracy.

\subsubsection{Optimizing  $\mathcal{L}_{tree}$} \label{sec412}
We design a pruning model to determine how to prune $T_g$ to minimize $\mathcal{L}_{tree}$.

The pruning model $h_\phi$ takes $T_g$ and $x$ (subscript omitted for simplicity) as input and outputs a vector $\mathbf{p}$, where each $p_i \in [0,1]$ represents the probability that the corresponding non-leaf node in $T_g$ should be pruned. Formally, 
\begin{align}
\mathop{\min}_T \mathcal{L}_{tree}(T) \\
\{T = T_g \circledast \mathbf{a}\} \subseteq T_g \label{eq9}\\
\mathbf{a} \sim \text{Bernoulli}(\mathbf{p}) \label{eq10}\\
\mathbf{p}=h_\phi(T_g,x) \label{eq11}
\end{align}
where pruning actions $\mathbf{a}$ is a binary vector, if $a_i=1$ then the $i^{th}$ node and its subtree in $T_g$ are removed.\footnote{Note that pruning is performed only on non-leaf nodes, so the size of $\mathbf{a}$ is $|\mathbf{a}|=|N_{T_g}| - |L_{T_g}|-1$ (excluding the root node).}
Each action $a_i$ is sampled from a Bernoulli distribution with parameter $p_i$. And $\mathbf{\theta}(T)$ in Eq. \ref{eq6} is treated as a known variable and thus omitted.

To optimize non-differentiable pruning actions, we utilize REINFORECE \cite{1992Simple} to maximize the expected performance of pruned trees. The objective is defined as:
\begin{equation}\label{eq12}
\begin{aligned}
\mathcal{L}_{tree}(T)&=-E_{T \sim \mathbf{p}}(R_T-R_{T_g})
\end{aligned}
\end{equation}
where $R_T$ is the evaluation metric of the tree $T$. We simply set $R_T=\text{XAUC}(y, \hat{y})-\text{MSE}(y,\hat y)$ (see Sec. \ref{sec513}). 

Eq. \ref{eq12} is self-critical, as it encourages sampling trees that outperform the global tree $T_g$, which also helps to reduce the variance of gradients.
The gradient is computed as follows:
\begin{equation} \label{eq13}
\begin{aligned}
& \nabla_\phi \mathcal{L}_{tree}(T) \\
&= -\nabla_\phi E_{T \sim \mathbf{p}}(R_T - R_{T_g}) \\
&= -E_{T \sim \mathbf{p}} \left[ (R_T - R_{T_g}) \nabla_\phi \log P(T \mid \mathbf{p}) \right] \\
&= -E_{T \sim \mathbf{p}} \left[ (R_T - R_{T_g}) \sum_{i} \nabla_\phi [a_i \log(p_i) + (1-a_i) \log(1-p_i)] \right] \\
\end{aligned}
\end{equation}

\subsection{Unbiased Conditional Learning} \label{sec42}
We further investigate how to achieve unbiased tree structure learning. In the original TPM framework, the tree structure introduces biased conditional learning: each classifier is trained only on samples falling within the interval defined by its parent node, but it may faces arbitrary ground truth watch times at inference.

Formally, the ideal cross entropy loss $\delta$ for a classifier is:
$\mathcal R=\frac{1}{|\mathcal{X}|} \sum_{x_i\in\mathcal{X}} \delta(o_i,\hat{o_i})$,
where $\mathcal{X}$ denotes the fully observed user-item pairs, $o_i\in\{0,1\}$ is the label $\mathbb{I}(y_i\in {n}_{\phi_{l_k}}(j)|x_i)$ and $\hat{o_i}=p(\hat y_i\in {n}_{\phi_{l_k}}(j)|x_i)$ is the prediction.
However, in TPM, to capture the conditional prior, $\hat o_i$ can only be trained for samples that $y_i\in {n}_{\phi_{l_k}}(j-1)$:
\begin{align}
\hat {\mathcal R}_{native}=\frac{1}{|x_i:y_i\in {n}_{\phi_{l_k}}(j-1)|} \sum_{x_i:y_i\in {n}_{\phi_{l_k}}(j-1)} \delta(o_i,\hat{o_i})
\end{align}
We call this the native estimator because it introduces selection bias, making $\hat {\mathcal R}_{native}$ a biased estimate of the true performance $\mathcal R$, i.e. $\mathbb{E}_{\mathbb{I}}[\hat {\mathcal R}_{native}] \neq \mathcal{R}$, where $\mathbb{I}$ is short for $\mathbb{I}(y_i\in {n}_{\phi_{l_k}}(j))$.

The key to handling this bias lies in understanding the process that generates the condition $y_i\in {n}_{\phi_{l_k}}(j-1)$. Fortunately, we can directly derive this generative process from Eq. \ref{eq1}:
\begin{equation}\label{eqn:generate_process} 
p(\hat y_i \in n_{\phi_{l_k}}(j-1) |  {x}_i) = \prod_{1 \leq l \leq j-2} p(\hat  y_i \in n_{\phi_{l_k}}(l) | \hat y_i \in n_{\phi_{l_k}}(l-1),  {x}_i)    
\end{equation}
where $p(\hat y_i \in n_{\phi_{l_k}}(j-1) |  {x}_i)$  is referred to as the propensity, modeling the probability that a sample belongs to a parent node. Then the Inverse-Propensity-Scoring (IPS) estimator \cite{thompson2012sampling,little2019statistical,imbens2015causal}  is,
\begin{equation}\label{eqn:debias-ips}
\hat {\mathcal R}_{IPS}=\frac{1}{|\mathcal{X}|} \sum_{x_i:y_i\in {n}_{\phi_{l_k}}(j-1)} \frac{\delta(o_i,\hat{o_i})}{p(\hat y_i \in n_{\phi_{l_k}}(j-1) |  {x}_i)}
\end{equation}
And the unbiased-ness is proved as follows:
\begin{equation}\label{eqn:unbiased-ness}
\begin{aligned}
\mathbb{E}_{\mathbb{I}}[\hat {\mathcal R}_{IPS}] 
&= \frac{1}{|\mathcal{X}|} \sum_{x_i\in \mathcal{X}} 
\mathbb{E}_{\mathbb{I}}[\frac{\delta(o_i,\hat{o_i})}{p(\hat y_i \in n_{\phi_{l_k}}(j-1) |  {x}_i)} \mathbb{I}] \\
&= \frac{1}{|\mathcal{X}|} \sum_{x_i\in \mathcal{X}} 
\frac{\delta(o_i,\hat{o_i})}{p(\hat y_i \in n_{\phi_{l_k}}(j-1) |  {x}_i)} \mathbb{E}_{\mathbb{I}}[\mathbb{I}] \\
&= \frac{1}{|\mathcal{X}|} \sum_{x_i\in \mathcal{X}} 
\delta(o_i,\hat{o_i}) = \mathcal{R} 
\end{aligned}
\end{equation}

\begin{algorithm}[t]
\caption{\textbf{PTPM}}
\label{alg1}
\begin{algorithmic}[1]
\STATE{Input:} Training data: $\{(x_i, y_i)\},\forall i$; a global tree: ${T_g}$;
\STATE{Output:} The classifiers of nodes $\{{f}_{\theta_j}\}$, the pruning model $h_\phi$;
\FOR{each batch}
\STATE Assign each sample $(x_i,y_i)$ to the leaf node $l_k$, and then assign the classifiers along the corresponding path $\phi_{l_k}$;
\STATE Compute $\mathcal{L}_{ce-UCL}$, the log-likelihood of $(x_i, y_i)$ belonging to path $\phi_{l_k}$, and $E(\hat y_i|x_i, T_g)$ and $Var(\hat y_i|x_i, T_g)$, as Eq. \ref{eq19}, \ref{eq4}.
\STATE Compute the pruning score $\mathbf{p}$, then sample pruning actions $\mathbf{a}$ and get a subtree $T$ via pruning on $T_g$ as Eq. \ref{eq9}, \ref{eq10}, \ref{eq11};
\STATE Compute the objective of $\mathcal{L}_{tree}$ as Eq. \ref{eq12};
\STATE Compute the final objective of $\mathcal{L}_{\text{PTPM}}$ as Eq. \ref{eq20};
\STATE Update $\{{f}_{\theta_j}\}$ and $h_\phi$ by minimizing $\mathcal{L}_{\text{PTPM}}$;
\ENDFOR
\end{algorithmic}
\end{algorithm}

The loss term $\mathcal{L}_{ce}$ in TPM can be upgraded as:
\begin{equation} \label{eq19}
\begin{aligned}
\mathcal{L}_{ce-UCL} &= -\log(p(\hat{y_i} \in l_k | x_i, T)) \\
&= -\sum_{1\leq j\leq d(l_k)} \frac{\log p(\hat y_i\in {n}_{\phi_{l_k}}(j) |\hat y_i\in {n}_{\phi_{l_k}}(j-1),x_i, T)}{p(\hat y_i \in n_{\phi_{l_k}}(j-1) |  {x}_i)}
\end{aligned}
\end{equation}

\subsection{Algorithm} In summary, the total objective function (Eq. \ref{eq20}) of PTPM consists of two parts, the TSL objective and the unbiased TPM objective. The training process is illustrated in Alg. \ref{alg1}.
\begin{equation} \label{eq20}
\begin{aligned}
\mathop{\min}_{T,\mathbf{\theta}} &\mathcal{L}_{\text{PTPM}} \\
&= \underbrace{\mathcal{L}_{\textbf{tree}}(T)}_{\text{TSL obj.}} + \underbrace{ \mathcal{L}_{\textbf{ce-UCL}}(\mathbf{\theta}) + \mathcal{L}_{reg}(\mathbf{\theta}) + \mathcal{L}_{var}(\mathbf{\theta})}_{\text{TPM obj.}}
\end{aligned}
\end{equation}

\subsection{Model Architecture}
\begin{figure}[htbp]
    \centering
    \includegraphics[width=0.55\columnwidth]{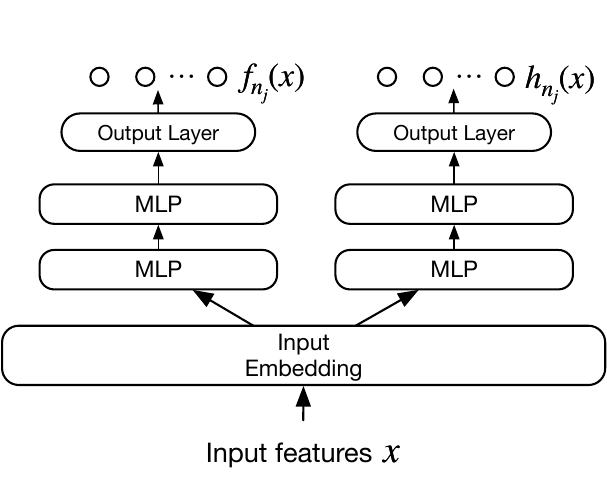}
    \caption{Architecture of the classifiers and the pruning model in PTPM. The outputs for classification and pruning tasks on node $n_j$ are $f_{n_j}(x)$ and $h_{n_j}(x)$, respectively.}
    \label{fig4}
\end{figure}

The architecture is presented in \textbf{Fig.} \ref{fig4}. Logically, each non-leaf node requires both a classifier and a pruning decision model, but this approach is too heavy for industrial RecSys. Hence, we adopt a single model that shares hidden layer parameters across tasks, with task-specific output layers for each node. Note that our framework can be easily extended to other advanced architectures.

\section{Experiments}
We conduct extensive experiments in both offline and online environments to demonstrate the effectiveness of PTPM.
  
\subsection{Experiment Setup}
\subsubsection{\textbf{Datasets.}} We conduct offline experiments on three public datasets: KuaiRec \cite{gao2022kuairec}, CIKM16\footnote{https://competitions.codalab.org/competitions/11161}, and WeChat21\footnote{https://github.com/WeChat-Big-Data-Challenge-2021/WeChat\_Big\_Data\_Challenge}. The KuaiRec dataset contains 7,176 users, 10,728 items, and 12,530,806 impressions. CIKM16, derived from the E-Commerce Search Challenge, aims to predict post-search dwell time and includes 87,934 users, 122,994 items, and 1,235,381 impressions. WeChat21 includes 20,000 users, 96,564 items, and 7.3 million impressions for predicting short video preferences.
\subsubsection{\textbf{Baselines}}
We select five state-of-the-art methods for comparison: WLR \cite{covington2016deep}, D2Q \cite{zhan2022deconfounding}, DML \cite{zhang2023leveraging}, CREAD \cite{sun2024cread}, and TPM \cite{lin2023tree}. Additionally, OR (Ordinal Regression \cite{crammer2001pranking}) is also chosen as a baseline. To ensure a fair comparison, the model architectures are kept consistent across approaches, with differences only in the output layers and their respective loss functions. As for PTPM, we set the depth of the global tree $d(T_g)=6$.

\begin{itemize}
    \item OR: Ordinal regression uses $K$ classifiers to predict whether the output exceeds each threshold.
    \item WLR: Watch time regression is framed as binary classification: clicked videos are positives, unclicked ones are negatives, and positive losses are weighted by watch time. In full-screen apps without click signals, we follow \cite{zhan2022deconfounding} and treat short plays as negatives.
    \item D2Q: This method groups samples by duration and trains the model to fit quantiles of watch time for each group.
    \item DML: Similar to D2Q, it considers the difficulty of estimating watch time across different intervals and sets coarser granularity at the beginning.
    \item CREAD: Building on OR, this method explores partitioning the watch time range via an offline search between equal-frequency and equal-width strategies. The resulting partitioning is then fixed for all samples.
    \item TPM: It organizes dependent classification tasks using a predefined full binary tree structure.
    
\end{itemize}

\subsubsection{\textbf{Metrics.}} \label{sec513}
Two evaluation metrics: MAE (Mean Absolute Error) and XAUC \cite{zhan2022deconfounding} are used.

\begin{itemize}
    \item MAE: This metric is a standard measure for assessing regression accuracy.
    $MAE = \frac{1}{N}\sum_{i=1}^N|\hat{y}_i-y|$   
    \item XAUC \cite{zhan2022deconfounding}: It evaluates whether the predicted order of two samples corresponds to their true watch time order.
    \begin{displaymath}
        \text{XAUC} = \frac{1}{N_{pair}} \sum_{(y_i, y_j)} \mathbb{I}\left((\hat{y}_i - \hat{y}_j)(y_i - y_j) > 0\right)
    \end{displaymath}
where $\mathbb{I}(\cdot)$ is the indicator function, and $N_{pair}$ is the total number of pairs.
\end{itemize}

\subsection{Offline Experiments}
\subsubsection{\textbf{Comparison with other methods}}
The results are listed in Table \ref{tb2}. PTPM achieves the best performance on all datasets. Compared to TPM, PTPM achieves superior performance, validating the effectiveness of tree structure learning. Furthermore, its comparison with CREAD further demonstrates PTPM's strong capability in learning to discretize.

\begin{table}[t]
\centering

\caption{Comparison between PTPM and other baselines. $*$ means significant improvements with $p < 0.05$.}
\resizebox{\columnwidth}{!}{%
\begin{tabular}{c|cc|cc|cc}
    \toprule
    \multirow{2}{*}{Approaches} & \multicolumn{2}{c|}{KuaiRec} & \multicolumn{2}{c|}{CIKM16} & \multicolumn{2}{c}{Wechat21} \\
    & MAE {\scriptsize $\downarrow$} & XAUC {\scriptsize $\uparrow$} & MAE {\scriptsize $\downarrow$} & XAUC {\scriptsize $\uparrow$} & MAE {\scriptsize $\downarrow$} & XAUC {\scriptsize $\uparrow$} \\
    \midrule
    WLR & 6.047 & 0.525 & 0.998 & 0.672 & 20.738 & 0.668 \\
    D2Q & 5.426 & 0.565 & 0.899 & 0.661 & 19.521 & 0.681 \\
    OR  & 5.321 & 0.558 & 0.918 & 0.664 & 19.876 & 0.679 \\
    DML & 5.150 & 0.563 & 0.892 & 0.663 & 19.489 & 0.682 \\
    CREAD & 5.230 & 0.586 & 0.910 & 0.669 & 19.534 & 0.682 \\
    TPM   & 4.741 & 0.599 & 0.884 & 0.676 & 19.355 & 0.688 \\
    PTPM & $\mathbf{4.536}^*$ & $\mathbf{0.622}^*$ & $\mathbf{0.810}^*$ & $\mathbf{0.693}^*$ & $\mathbf{19.210}^*$ & $\mathbf{0.710}^*$ \\
    \bottomrule
\end{tabular}
}
\label{tb2}
\end{table}

\subsubsection{\textbf{Ablation Studies.}}

We first analyze the contribution of different modules to the final results. As shown in Table \ref{tb3}, the entire method outperforms all the variants. The comparison between PTPM with/without TSL shows the effectiveness of tree structure learning. Compared to PTPM(w/o UCL), the UCL module achieves consistent improvements which demonstrates that the debiasing of SSB improves the generalization of classifiers.

\textit{Efficiency of tree learning.} As shown in \textbf{Fig. \ref{fig5}}, we first confirm that the number of parameters and computational costs of PTPM is almost equal to those of TPM. Moreover, the TSL method converges significantly faster and achieves better final performance. This improvement is attributed to the proposed pruning paradigm, which avoids costly inner optimization and enables joint training of the tree structure and classifiers. 

\begin{table}[t]
\centering
\caption{PTPM with different components}
\resizebox{\columnwidth}{!}{%
\begin{tabular}{c|cc|cc|cc}
    \toprule
    \multirow{2}{*}{Approaches} & \multicolumn{2}{c|}{KuaiRec} & \multicolumn{2}{c|}{CIKM16} & \multicolumn{2}{c}{Wechat21} \\
    & MAE {\scriptsize $\downarrow$} & XAUC {\scriptsize $\uparrow$} & MAE {\scriptsize $\downarrow$} & XAUC {\scriptsize $\uparrow$} & MAE {\scriptsize $\downarrow$} & XAUC {\scriptsize $\uparrow$} \\
    \midrule
    PTPM & $\mathbf{4.536}$ & $\mathbf{0.622}$ & $\mathbf{0.810}$ & $\mathbf{0.693}$ & $\mathbf{19.210}$ & $\mathbf{0.710}$ \\
    PTPM w.t.o. TSL  & 4.650 & 0.605 & 0.870 & 0.680 & 19.330 & 0.691 \\
    PTPM w.t.o. UCL  & 4.600 & 0.615 & 0.860 & 0.686 & 19.300 & 0.697 \\
    TPM                & 4.741 & 0.599 & 0.884 & 0.676 & 19.355 & 0.688 \\
    \bottomrule
\end{tabular}
}
\label{tb3}
\end{table}

\begin{figure}[htbp]
    \centering
        \begin{subfigure}[b]{0.45\columnwidth}
        \centering
        \begin{tikzpicture}[scale=0.6]
            \begin{axis}[
                width=6cm, height=6cm,
                ybar=0.4cm,
                bar width=0.3cm,
                symbolic x coords={\#Parameters, \#Computation(flops)},
                xtick=data,
                ymin=20, ymax=75,
                ylabel={Value (M)},
                nodes near coords={
                    \pgfmathprintnumber{\pgfplotspointmeta} 
                },
                every node near coord/.append style={
                    font=\small,
                    anchor=south,
                    yshift=2pt
                },
                legend style={
                    at={(0.02,0.98)},
                    anchor=north west,
                    draw=black,
                    legend columns=1,
                    /tikz/every even column/.append style={column sep=10pt}
                },
                enlarge x limits=0.25
            ]
            \addplot+[
                ybar,
                xshift=1pt
            ] coordinates {(\#Parameters, 32.274079) (\#Computation(flops), 64.547659)};

            \addplot+[
                ybar,
                xshift=-1pt
            ] coordinates {(\#Parameters, 32.548158) (\#Computation(flops), 65.095311)};
            
            \legend{TPM, PTPM}
            \end{axis}
        \end{tikzpicture}
    \end{subfigure}
    \hfill
    \begin{subfigure}[b]{0.45\columnwidth}
        \centering
        \begin{tikzpicture} [scale=0.6]
        \begin{axis}[
            width=6cm, height=5.5cm,
            xmin=-1000, xmax=45000,
            ymin=0.45, ymax=0.65,
            xlabel={\# Search steps},
            ylabel={Max XAUC},
            grid=major,
            legend pos=south east, 
            legend style={nodes={scale=0.8, transform shape}},
            xtick={0,5000,...,50000} 
        ]
        
        \addplot[
            color=red,
            mark=*,
            mark size=1pt
        ]
    table[col sep=comma] {data.txt};
        \addlegendentry{TSL}
    
        \addplot[
            color=blue,
            mark=*,
            mark size=0.7pt 
        ]
    table[col sep=comma] {data2.txt};

        \addlegendentry{Random Search}

        \addplot[
            color=blue,
            line width=1.1pt,
            dashed
        ]
        coordinates {
            (0,0.60) (49999,0.60)
        };
        \addlegendentry{Global Tree: $T_g$}
        
        \end{axis}
    \end{tikzpicture}
    \end{subfigure}
    \caption{\textit{Left}: Model size and flops; \textit{Right}: Efficiency of learning tree structure on KuaiRec dataset.}
    \label{fig5}
\end{figure}
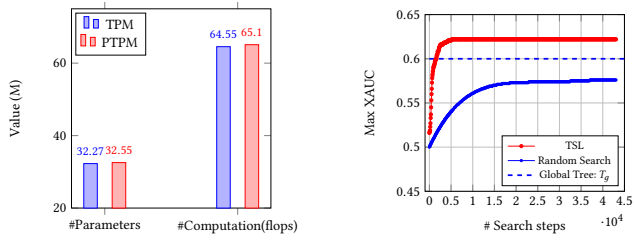

\begin{figure} [htbp]
    \centering
    
    \begin{subfigure}[t]{0.45\columnwidth}
        \centering
        \begin{tikzpicture}[scale=0.6]
    \begin{axis}[
        width=6cm,
        height=5.5cm,
        xmin=1, xmax=32,
        ymin=0.3, ymax=2,
        xlabel={Leaf node $l_k$ in $T_g$ with depth 6},
        ylabel={Ratio of prior-to-posterior},
        xtick={1,4,...,32},
        ytick={0.5,1,...,2},
        grid=major,
        legend pos=north west
    ]
    
    \addplot[
        color=blue,
        thick,
        dashed
    ]
    coordinates {
        (1,1.0) (32,1.0)
    };
    \addlegendentry{Ideal: $y=1.0$}

    \addplot[
        color=red,
        thick
    ]
    coordinates {
        (1,0.69045944) (2,0.75246923) (3,0.86720101) (4,0.91301724)
        (5,0.92551755) (6,0.92492268) (7,0.93044535) (8,0.95059223)
        (9,0.94203586) (10,0.92735824) (11,0.98311099) (12,0.95333273)
        (13,0.96230704) (14,1.00506051) (15,0.97727906) (16,0.98046364)
        (17,1.0009214) (18,1.00969278) (19,1.00978174) (20,1.00363096)
        (21,0.9948264) (22,1.00829369) (23,0.98189825) (24,1.02169727)
        (25,0.98352912) (26,1.02271377) (27,1.02534396) (28,1.05173107)
        (29,1.08408889) (30,1.19650071) (31,1.31766904) (32,1.81267311)
    };
    \addlegendentry{TPM}

    \addplot[
    color={rgb:red,2;green,1;blue,0.5},
        line width=1.2pt
    ]
    coordinates {
        (1,0.910370166) (2,0.95630712) (3,0.95928794) (4,0.95502807)
        (5,0.96602669) (6,0.96194718) (7,0.96371934) (8,0.99805637)
        (9,1.00983355) (10,1.01172135) (11,1.02077884) (12,1.03127859)
        (13,1.04265241) (14,1.01356826) (15,1.02432868) (16,1.02038169)
        (17,1.01823152) (18,1.02780251) (19,1.03141938) (20,1.0309423)
        (21,1.00087905) (22,1.03311452) (23,1.03038954) (24,1.02716404)
        (25,1.03118427) (26,1.03051732) (27,0.98828617) (28,1.03474124)
        (29,1.05545912) (30,1.03133867) (31,1.05950683) (32,1.13849467)
    };
    \addlegendentry{TPM+UCL}
    \end{axis}
\end{tikzpicture}  

    \end{subfigure}
    \hfill
    \begin{subfigure}[t]{0.48\columnwidth}
        \centering
        \begin{tikzpicture} [scale=0.6]
        \begin{axis}[
            width=5.5cm, 
            height=5.5cm, 
            ymin=0.55, ymax=0.66, 
            ybar, 
            symbolic x coords={TPM, TPM+UCL}, 
            xtick=data, 
            nodes near coords, 
            bar width=0.4cm, 
            ylabel={Avg. AUC of Classifiers}, 
            enlarge x limits=0.75, 
            yminorticks=false
        ]
        \addplot coordinates {(TPM,0.594) (TPM+UCL,0.623)};
        \end{axis}
        \end{tikzpicture}  
    \end{subfigure}

    \caption{Analysis of Unbiased Conditional Learning}
    \label{fig6}
\end{figure}
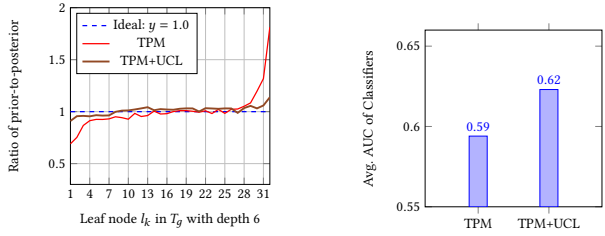

\textit{Analysis of Unbiased Conditional Learning (UCL).} We employ two metrics on KuaiRec dataset to analyze the calibration effect on SSB. The one is the ratio of prior-to-posterior, $\frac{\sum_i p(\hat y_i \in l_k|x_i)}{\sum_i \mathbb{I}(y_i\in l_k |x_i)}$, which evaluates the accuracy of the predicted probability that watch time belongs to a leaf node $l_k$ and it is expected to 1.0; the other is the average AUC of the classifiers. 
As shown in \textbf{Fig. \ref{fig6}}, the original TPM underestimates in left-side intervals and overestimates in right-side intervals. The UCL adjusts the estimates closer to 1.0 and significantly improves the  AUC of the classifiers.

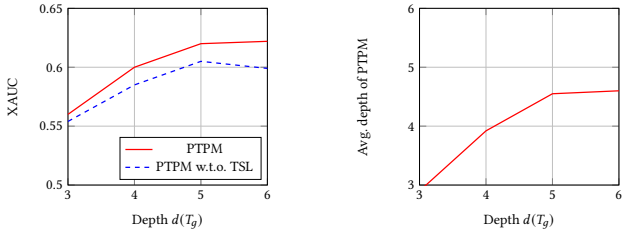
\begin{figure} [htbp]
    \centering
    
    \begin{subfigure}[t]{0.45\columnwidth}
        \centering
        \begin{tikzpicture}[scale=0.6]
    \begin{axis}[
        width=6cm,
        height=5.5cm,
        xmin=3, xmax=6,
        ymin=0.5, ymax=0.65,
        xlabel={Depth $d(T_g)$},
        ylabel={XAUC},
        xtick={3,4,...,8},
        grid=major,
        legend pos=south east
    ]
    
    \addplot[
        color=red,
        thick
    ]
    coordinates {
        (3,0.56) (4,0.60) (5,0.62) (6,0.622) 
    };
    \addlegendentry{PTPM}

    \addplot[
        color=blue,
        thick,
        dashed
    ]
    coordinates {
        (3,0.554) (4,0.585) (5,0.605) (6,0.599) 
    };
    \addlegendentry{PTPM w.t.o. TSL}

    \end{axis}
\end{tikzpicture}  

    \end{subfigure}
    \hfill
        \begin{subfigure}[t]{0.45\columnwidth}
        \centering
        \begin{tikzpicture}[scale=0.6]
    \begin{axis}[
        width=6cm,
        height=5.5cm,
        xmin=3, xmax=6,
        ymin=3, ymax=6,
        xlabel={Depth $d(T_g)$},
        ylabel={Avg. depth of PTPM},
        xtick={3,4,...,8},
        ytick={3,4,...,8},
        grid=major,
        legend pos=south east
    ]
    
    \addplot[
        color=red,
        thick
    ]
    coordinates {
        (3,2.9) (4,3.92) (5,4.55) (6,4.60) 
    };

    \end{axis}
\end{tikzpicture}  

    \end{subfigure}

    \caption{Analysis of global tree depth $d(T_g)$ on KuaiRec}
    \label{fig7}
\end{figure}
\textit{Parameter sensitivity (Fig. \ref{fig7}).} We study the sensitivity to the global tree depth $d(T_g)$. PTPM consistently outperforms PTPM (w/o TSL) across a range of tree depths, and its performance improves with increasing depth. This suggests that deeper trees provide a larger structural search space, enabling TSL to discover more effective structures. Meanwhile, the performance improvement of PTPM becomes smaller as the depth increases. And the average depth of learned trees converges to around 4.6 when $d(T_g) \geq 6$. This suggests that $d(T_g) = 6$ is a cost-effective choice.

\begin{table}[htbp]
  \caption{Online A/B testing results on Kuaishou. Statistically significant results (p-value < 0.05) are highlighted in bold.}
  \label{tb4}
  \begin{tabular}{@{}p{1.5cm}p{0.8cm}|p{0.9cm}p{1.3cm}@{}}
    \hline
    \multicolumn{2}{l|}{Main metric}      & \multicolumn{2}{c}{Constraints} \\ \hline
    \multicolumn{2}{l|}{Watch time}       & Latency & CPU time \\ \hline
    \multicolumn{2}{l|}{\textbf{+0.428\%}} & 0.006\% & 0.052\%  \\ \hline
  \end{tabular}
\end{table}

\subsection{Online Experiments}
\subsubsection{\textbf{Experiment Setup}}
We conduct A/B experiments on the main page (i.e., the Featured tab) of Kuaishou App. The traffic is split into ten buckets uniformly, and we attribute 20\% traffic to our PTPM while another 20\% is assigned to the online baseline TPM. Kuaishou serves over 400 million users daily and the results are very convincing. For PTPM, the global tree is designed as a full binary tree with $d(T_g)=6$. For online baseline TPM, its tree structure is the same with the global tree for fair comparison.
The predicted watch time is a positive component used in ranking stage. In the experimental group, the prediction score from TPM was replaced with that from PTPM.
\subsubsection{\textbf{Experiment Results}}
We launch the experiments on the system for a whole week and list the results in Table \ref{tb4}. Watch time is the primary metric, while latency and CPU time are constrained metrics.
Notice that PTPM outperforms the baseline in watch time metric. Meanwhile, the latency is nearly identical to the baseline and the additional computation cost is minimal and acceptable, demonstrating that PTPM is very efficient.

\section{Conclusion}
This paper presents PTPM, a personalized extension of the Tree-based Progressive Regression Model (TPM), to address the challenge of watch time prediction in video recommendation systems. Unlike the original TPM, which employs a fixed full binary tree structure, PTPM dynamically selects optimal tree structures for different samples under a unified objective. This adaptability allows the model to better capture the diversity of user behaviors and video characteristics, leading to improved prediction performance. We also tackle a significant issue in TPM: sample selection bias due to tree-based conditional modeling. To mitigate this, we employ a plug-and-play approach that reduces bias without altering the model's core structure. Extensive experiments on both offline datasets and online environments demonstrate PTPM's superiority over existing methods.

In summary, PTPM enhances predictive accuracy and offers a more flexible and robust solution for watch time prediction in video recommender systems. Future work will focus on extending PTPM to handle more complex user-item interactions and exploring its applications in dynamic reasoning.

\section{GenAI Usage Disclosure}
No generative AI tools were used in any part of this work, including the design of experiments, code implementation, data analysis, and manuscript writing.

\bibliographystyle{ACM-Reference-Format}
\bibliography{sample-base}

\end{document}